\documentclass[11pt]{article}
\usepackage[dvips]{graphicx}
\usepackage{authblk}
\usepackage{amsmath,amsfonts,amsthm,amssymb,bm}
\usepackage{subfigure}
\usepackage{enumerate}
\usepackage{rotating}
\usepackage{multirow}
\usepackage{url}
\usepackage{enumitem}
\usepackage{slashbox}
\usepackage{cases}
\usepackage{fullpage}
\begin{document}
\title{Gossip-based Search in Multipeer Communication Networks}
\author[$\dagger$]{Eva Jaho}
\author[*]{Ioannis Koukoutsidis}
\author[$\ddagger$]{Siyu Tang}
\author[$\dagger$]{\\Ioannis Stavrakakis}
\author[$\ddagger$]{Piet Van Mieghem}
\affil[$\dagger$]{National \& Kapodistrian University of
Athens\\Dept. Informatics and Telecommunications\\Ilissia, 157 84
Athens, Greece\\ E-mail: \{ejaho,ioannis\}@di.uoa.gr}

\affil[ ]{}
\affil[*]{University of Peloponnese, Dept. Telecommunications Science and Technology\\\qquad
End of Karaiskaki St., 22100, Tripolis, Greece\\E-mail: gkouk@uop.gr}
\affil[ ]{}
\affil[$\ddagger$]{Delft University of Technology\\2600 GA Delft,
The Netherlands\\ E-mail: \{S.Tang,P.F.A.VanMieghem\}@tudelft.nl} 
\date{June 21, 2009}
\maketitle
\begin{abstract}
We study a gossip-based algorithm for searching data objects in a multipeer communication
network. All of the nodes in the network are able to communicate with each other.
There exists an initiator node that starts a round of searches by randomly querying one or
more of its neighbours for a desired object. The queried nodes can also be activated
and look for the object. We examine several behavioural patterns of nodes with respect to their
willingness to cooperate in the search. We derive mathematical models for the search process
based on the balls and bins model, as well as known approximations for the rumour-spreading problem.
All models are validated with simulations. We also evaluate the performance
of the algorithm and examine the impact of search parameters.
\end{abstract}

\section{Introduction}
The term `gossiping algorithm' encompasses any communication algorithm where messages between two nodes
are exchanged opportunistically, with the intervention of other nodes that act as betweeners or forwarders
of the message. It is inspired from the social sciences, in the same way as epidemic protocols where
inspired from the spreading of infectuous diseases \cite{Eugster04}. These two communication paradigms
are very similar, with differences focusing on the different ways that gossiping nodes and
infected nodes could behave: gossiping nodes adopt human-like characteristics, while the behaviour of
infected nodes is governed by the dynamics of the virus or disease.

Gossiping algorithms are suitable for communication in distributed systems, such as ad-hoc networks and generally
systems with peer-to-peer or peer-to-multipeer communication.
The latter communication paradigm is followed here, where a node can communicate with multiple peers,
usually maintaining a short-time connection with each one.
Attractive characteristics of gossiping algorithms include simplicity, scalability and robustness to failures,
as well as a speed of dissemination that is easily configurable.
Gossiping can be identified with the spreading of rumours in a network, the
dymanics of which are investigated in \cite{Pittel87,Nekovee08}.
Additionally, gossiping protocols have been used for the computation of aggregate network quantities, such
as sums, averages, or quantiles of certain node values \cite{Kempe03}.

In all of the above referenced works, gossiping is typically used for the dissemination of information,
and performance metrics are oriented towards measuring the efficiency of information dissemination.
In this paper, we model a specific gossip-based algorithm that aims at finding files (or data
objects, in general) in nodes in a distributed network. The algorithm employs sequentially-generated
parallel search procedures, in the following manner:
We assume there exists a file in the network that may be located in different nodes.
An {\em initiator} node is interested in this file and starts a round of searches to find it, by randomly querying
one or more of its peers (neighbours). The queried nodes can also be {\em activated}
and look for the object. The search is considered successful when at least one copy of the file is found.

Apart from the initiator, the other nodes that assist the search can have different behavioural patterns.
We distinguish between {\em cooperative} and {\em non-cooperative} nodes. Nodes in the first category always become
active when queried, and generate themselves query messages in subsequent rounds.
Non-cooperative nodes on the other hand are unwilling to participate in the search process themselves.
We also consider {\em stifler} nodes; the term is borrowed from \cite{Nekovee08} and signifies nodes that were previously active,
but from a certain point on lose interest in the dissemination of the query, and thus cease to participate in the search.
Hence, it is a special case of cooperation. To avoid confusion in the paper, non-stifler nodes that are non-cooperative
are also referred to as {\em plain non-cooperative} nodes.
In all the above cases cooperation is considered only with respect to participating in the search;
if a node has the file it always returns it.

We also derive different versions of the algorithm based on the level of knowledge that each node has
about the progress of search. We consider two extremes: at the one, each node has no knowledge whatsoever
about the number or identities of nodes that have been previously queried in the network.
At the other extreme, each node has complete knowledge about these facts and avoids sending messages
to previously queried nodes at subsequent rounds. We call these cases {\em blind search} and {\em smart search},
respectively. 

We mathematically model the blind search process based on a known approximation for the rumour spread\-ing problem \cite{Pittel87},
which we extend here. The smart search process is modeled using a combinatorial approach, based on a generalization
of the balls and bins model \cite{Feller68}. Based on these models, we are able to evaluate the performance
of the search, as well as the impact of search parameters. The latter are the number of queried neighbours by a node
and the number of copies of the file in the network. By changing these parameters, one can easily configure the
speed and efficiency of the search, as will be shown later.

Both the blind and smart versions of the gossiping algorithm have been modeled exactly 
in \cite{Tang09}. Both information dissemination and search are investigated in that paper,
while here we are focusing on the search process.
The main algorithmic difference in \cite{Tang09} is that even the nodes that have the file can be non-cooperative.
In this paper, we want to focus only on the effect that cooperation has in the forwarding of the message: only the intermediate nodes forwarding a query can be non-cooperative.
In addition, the approximative model presented here for the blind search process is shown to be computationally simpler,
while maintaining good accuracy. Finally, in this paper we present the stifling behavioral pattern, which
is not included in \cite{Tang09}.

The paper is structured as follows. In Section~\ref{sec:algo}, the gossip-based search algorithm is described
in more detail, and application scenarios that justify the study of the blind and smart search algorithms are discussed.
In Sections \ref{sec:blind} and \ref{sec:smart} we present the mathematical modeling of the blind and smart search algorithms, respectively.
The modeling in these sections covers cooperative and plain non-cooperative nodes. The stifling behavioural
pattern is analysed in Section~\ref{sec:stiflers}. In Section~\ref{sec:scaling}, we present results for the performance of the blind search algorithm for very large numbers of nodes, and derive useful scaling laws. The major conclusions from this work and 
issues for future research are presented in Section~\ref{sec:conclusions}.
\section{Gossip-based search algorithm}\label{sec:algo}
A model of the network in the form of a complete graph is considered. There is an initiator node $\mathcal{I}$,
and $N-1$ other nodes in the graph. There is a file $f$ located in $m$ of the other nodes of the graph ($m\leq N-1$) that
the initiator wants to find. The initiator starts a search by randomly
querying a subset of its neighbors of size $k$ ($k\leq N-1$), with equal probabilities.

If a queried node has the object then it returns it, and the query is successful. Otherwise the queried nodes
-- depending on being cooperative or not -- may begin to search themselves by forwarding the query to their neighbours.
Cooperative nodes which are queried become ``active'' and participate in the
search.
The process of the search is modeled in steps or rounds, where at each round
all active nodes simultaneously query their neighbors,
hence activating new nodes. The algorithm continues for several rounds where at each
step, active nodes randomly query some of their
neighboring nodes, until the file $f$ is found.

We consider two search scenarios:
\begin{itemize}
\item[-] {\em Blind search}: An active node searches ``blindly'' at each round, possibly querying nodes that have been queried before. This approach can model devices with small
computational capabilities, that cannot keep a log of queried nodes, or cases
where the identities of the devices are not known. It is equally appropriate to model
situations with random encounters between nodes. For instance, a
number of mobility models have exponential meeting times
between mobile nodes (such as the Random Walk, Random Waypoint and
Random Direction models, as well as more realistic, synthetic models
based on these \cite{Spyropoulos08}). In our model, the time until a node is queried
approaches a geometric distribution, which is the discrete time analog to an exponential distribution.
\item[-] {\em Smart search}: An active node searches ``smartly'' at each round,
by avoiding nodes that have been queried before either by itself or by other
nodes. This demands the knowledge of the identities of all queried nodes,
and has a larger overhead compared to the blind search case.
We do not define the exact algorithm by which the identities of all queried nodes
are made known to an active node. We only assume that this knowledge can be obtained
at a cost that is small compared to the cost of searching, and use this case mainly
as a reference for the efficiency of the blind search algorithm.
It is evident that smart search corresponds to the fastest version of the algorithm. Although it can be hard and costly to implement,
there can exist schemes that can approximate its performance. For example, a low-cost algorithm that
could approximate smart search can be based on the routine that, at each peer-to-peer communication, nodes
exchange the lists of peers they have queried.
\end{itemize}
\section{Approximate blind search model with cooperative or non-\\cooperative nodes}\label{sec:blind}
\label{approx_blind_search}
Each node that receives a search query will
cooperate to forward the query with probability $c$ ($0\leq c\leq 1$).
If several active nodes query the same node, the latter node decides whether 
to be cooperative or not by a single Bernoulli trial. 
We do not consider that the node performs several independent trials, one for each query.

We consider a sequence of steps (or rounds) $r=1,2,\dots$ until the
file is found. If at step $r$ there are $\hat{A}(r)$ active nodes then,
provided the file is not yet found, the probability of finding it at the
$r$th step, $S(r)$, is:
\begin{equation}\label{incl-excl}
S(r)=1-\left(1-p_s\right)^{\hat{A}(r)}\;,
\end{equation}
where $p_s$ is the probability that a single search (consisting of $k$
different random queries) succeeds.

To find $p_{s}$, notice that the problem is equivalent to the
one where, in a set of $N-1$ nodes, there are $m$ marked nodes and we
randomly select a group of $k$ nodes. We want to find the
probability that at least one marked node is selected. The probability
that our selection returns exactly $u$ marked nodes ($u\leq
\min({m,k})$) is
\begin{displaymath}
p_u=\frac{\binom{m}{u}\binom{N-1-m}{k-u}}{\binom{N-1}{k}}\;.
\end{displaymath}
Indeed, the marked nodes can be chosen in
$\binom{m}{u}$ different ways, the unmarked ones in
$\binom{N-1-m}{k-u}$ ways, and the total number of ways to select $k$ nodes is $\binom{N-1}{k}$.
Further, $p_{s}=1-p_0$, therefore
\begin{equation}\label{single-search}
p_s=1-\frac{\binom{N-1-m}{k}}{\binom{N-1}{k}}\;.
\end{equation}

The probability of finding the file $f$ at the $r$th
step is,
\begin{equation}\label{prob_find_bs}
p(r)=S(r)\prod_{i=1}^{r-1}(1-S(i))\;.
\end{equation}
This formula is an approximation because it implicitly assumes that
each round is independent of the other.

A deterministic approximation $\hat{A}(r)$ for the number of active nodes in each
round can be found using the method presented in \cite{Pittel87},
which is extended to $k$ neighbours that are cooperative with probability $c$. Consider the process $\{I(r),\, r\geq 1\}$ of the number of inactive nodes in
each round. Given that at round $r$ there are $i(r)$ inactive nodes
and ${A}(r)$ active ones, the mean number of inactive nodes at
round $r+1$ will be
\begin{align}
E[I(r+1)]=& i(r)\left[\left(1-\frac{k}{N-1}\right)^{{A}(r)}+\left[1-\left(1-\frac{k}{N-1}\right)^{{A}(r)}\right](1-c)\right]\nonumber\\
=& i(r)\left[(1-c)+c\left(1-\frac{k}{N-1}\right)^{{A}(r)}\right]\nonumber\;.
\end{align}

For fixed $k$ and large $N$, we use a second order expansion of $(1-\frac{k}{N-1})^{A(r)}$, so that $(1-\frac{k}{N-1})^{{A}(r)}\approx
e^{-{A}(r)(\frac{k}{N-1}+\frac{k^2}{2(N-1)^2})}$, and
\begin{align}
E[I(r+1)]=i(r)\left[(1-c)+ce^{-{A}(r)(\frac{k}{N-1}+\frac{k^2}{2(N-1)^2})}\right]\nonumber\;.
\end{align}
From this, by assuming $I(r)=i(r)\:\forall r$, we derive the deterministic approximation
\begin{align}
I(r+1)=I(r)\left[(1-c)+ce^{-{A}(r)(\frac{k}{N-1}+\frac{k^2}{2(N-1)^2})}\right]\;.
\end{align}
Using that ${A}(r)=N-I(r)$, we finally obtain the recursion:
\begin{equation}\label{approx_recursion}
\begin{split}
{A}(r+1)=Nc+A(r)(1-c)-(N-A(r))ce^{-{A}(r)(\frac{k}{N-1}+\frac{k^2}{2(N-1)^2})}\;,
\end{split}
\end{equation}
with $A(1)=1$.
Since ${A}(r)$ is not an integer in general, we round it to the
nearest integer, which we denote by
$\hat{A}(r)=[{A}(r)]$.

Following a similar approach as in \cite{Pittel87}, it can be shown that
the distribution of $I(r+1)$, given $i(r)$ is indeed concentrated sharply
around $i(r)[(1-c)+c\exp({-{A}(r)(\frac{k}{N-1}}$ 
$+\frac{k^2}{2(N-1)^2}))]$, and that the approximation
becomes more accurate as $k/N\to 0$.

From (\ref{prob_find_bs}), we have constructed an approximate distribution for the
number of steps until the file is found. We can then derive the mean number of steps
until the file is found and the mean number of nodes $A$ involved in the search
(activated nodes):
\begin{align}
E[r]= &\sum_{r=1}^{\infty} r p(r),\label{mean_rounds}\\
E[A]= &\sum_{r=1}^{\infty} \hat{A}(r+1) p(r)\;.
\end{align}
(Notice that $\hat{A}(r+1)$ nodes will be activated approximately at round
$r$.)

For numerical calculations, as an upper bound on
the support of $r$ we take
\begin{equation}\label{eps}
r_{max}=\min\{r: \prod_{i=1}^{r-1}(1-S(i))<\epsilon\}\;,
\end{equation}
where $\epsilon$ is a number close to zero.
\theoremstyle{remark}
\newtheorem{rem}{Remark}
\begin{rem}
In \cite{Tang09}, we have derived an exact model for a slightly
different version of the blind search algorithm. 
Generally, the exact approach
for modeling the search algorithm requires the calculation of the $N\times N$
transition matrix $Q_{[ij]}$, where the $(i,j)$-th value is the probability of going
from $i$ to $j$ active nodes in one round. Then $Q^r(1,i)$ denotes the probability that
there are $i$ active nodes in $r$ rounds.

The probability of finding the file in $r$ rounds, denoted here by $B(r)$, can be calculated as
\begin{equation}\label{prob_findby_exact}
B(r)=\sum_{i=1}^{N}\left[1-\frac{\binom{N-i}{m}}{\binom{N-1}{m}}\right]Q^r(1,i)\;.
\end{equation}
This is a probability distribution, therefore the probability of finding the file exactly
at round $r$ is given by:
\begin{equation}\label{prob_find_exact}
B(r)-B(r-1)\;.
\end{equation}
In the Appendix, we compare the complexity of the two models and examine the accuracy of the approximate one.
It is shown that the reduction in computational cost is of the order of $O(N^2)$, while
the relative accuracy of the approximation is higher than 95\% in the majority of cases
(the comparison holds when $c=1$).
\end{rem}

\begin{figure}[!htb]
\centering
\subfigure[]{
\includegraphics[scale=0.75]{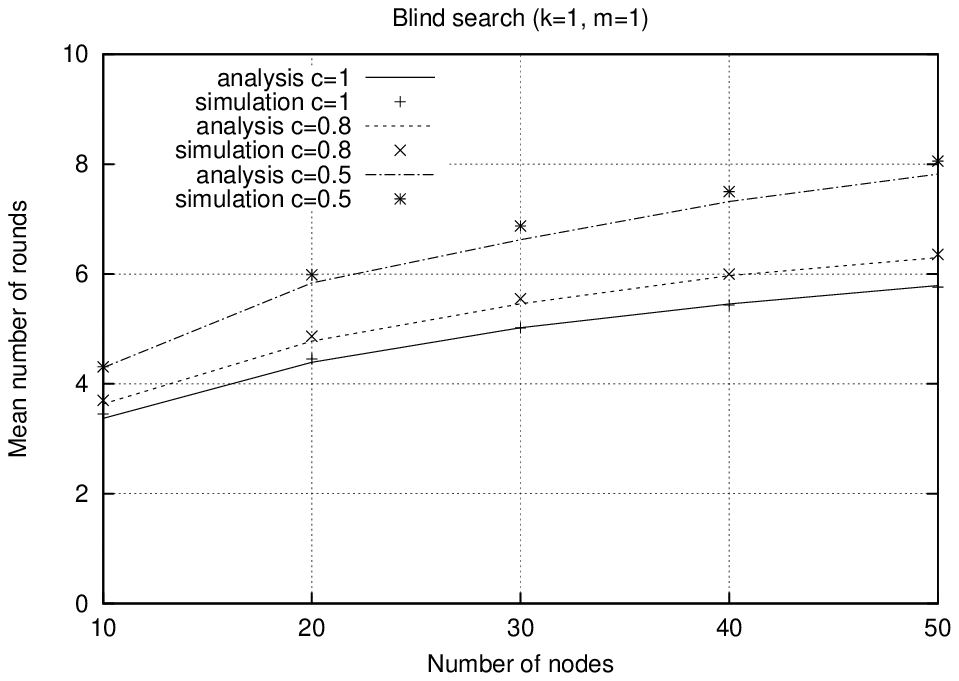}
\label{fig:subfig1}} %
\hspace{0pt}%
\subfigure[]{
\includegraphics[scale=0.75]{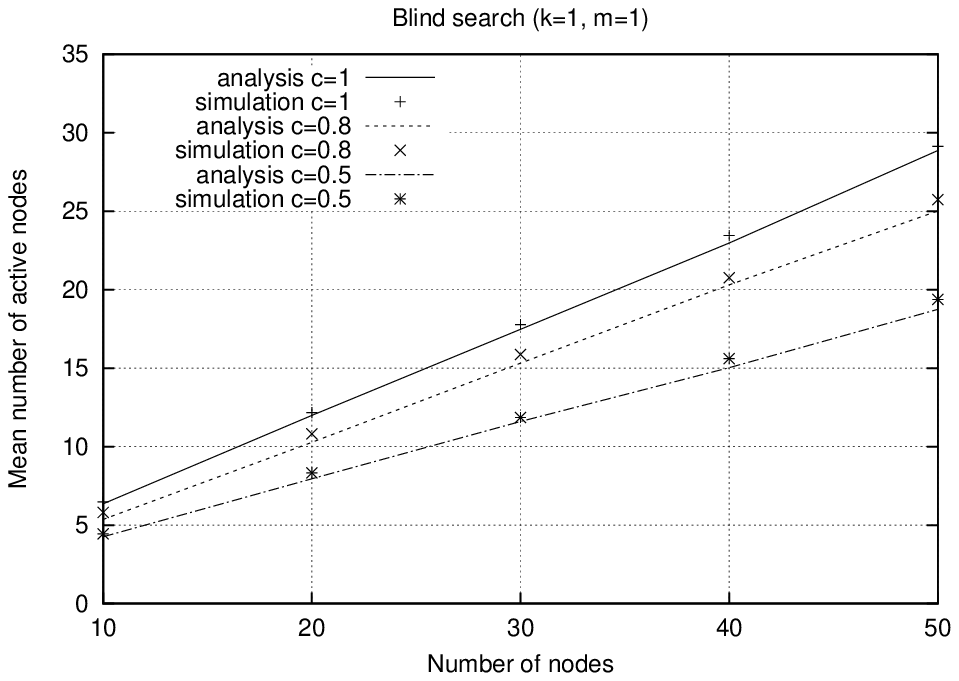}
\label{fig:subfig2}}\\%
\vspace{6pt}%
\subfigure[]{
\includegraphics[scale=0.75]{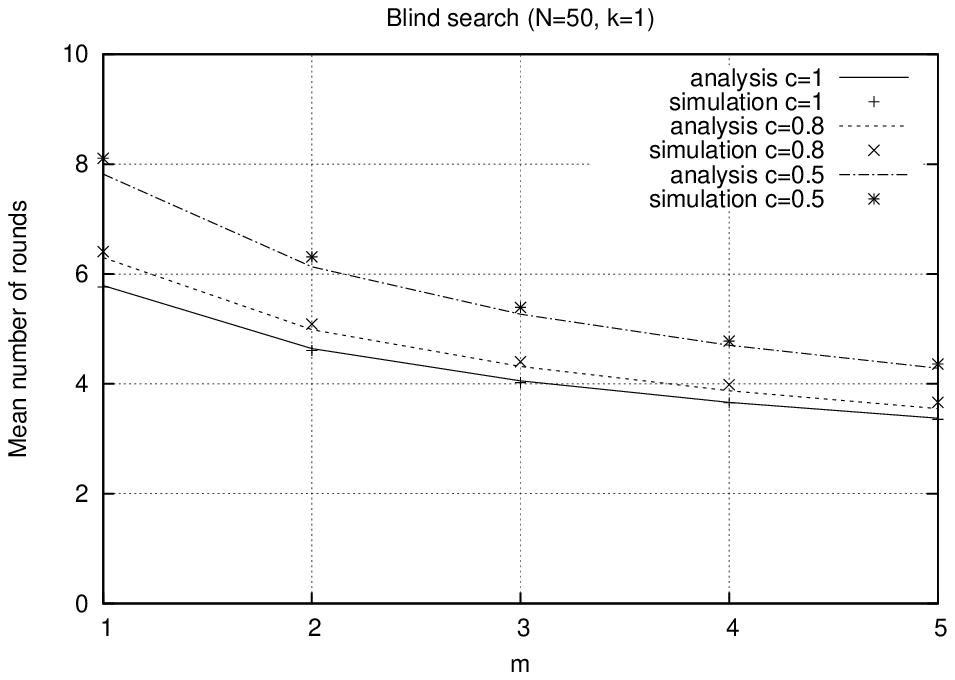}
\label{fig:subfig3}}%
\hspace{0pt}%
\subfigure[]{
\includegraphics[scale=0.75]{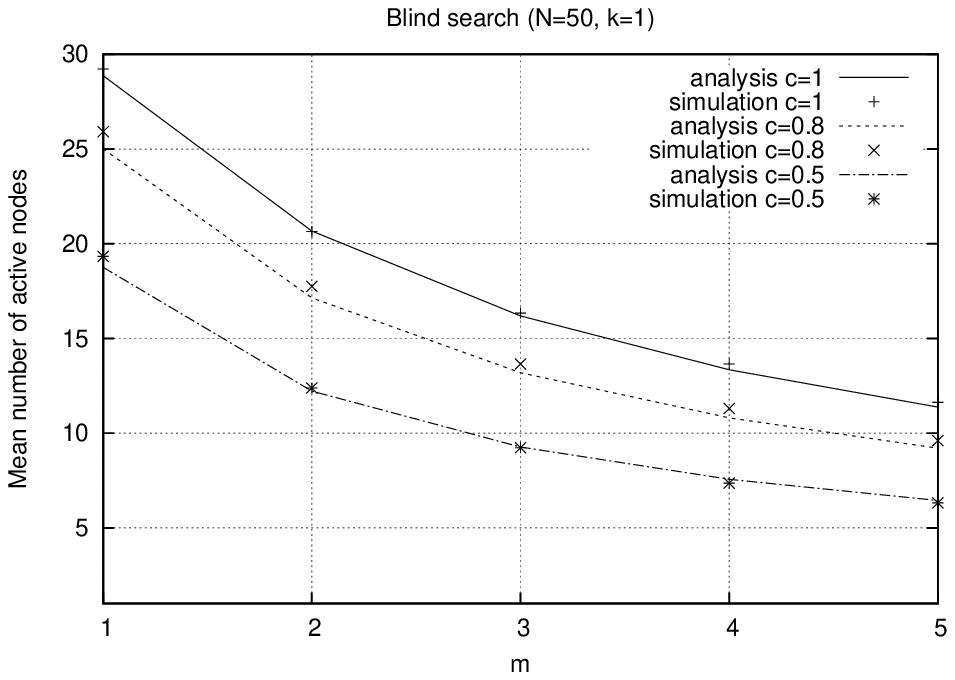}
\label{fig:subfig4}}\\%
\vspace{6pt}%
\subfigure[]{
\includegraphics[scale=0.75]{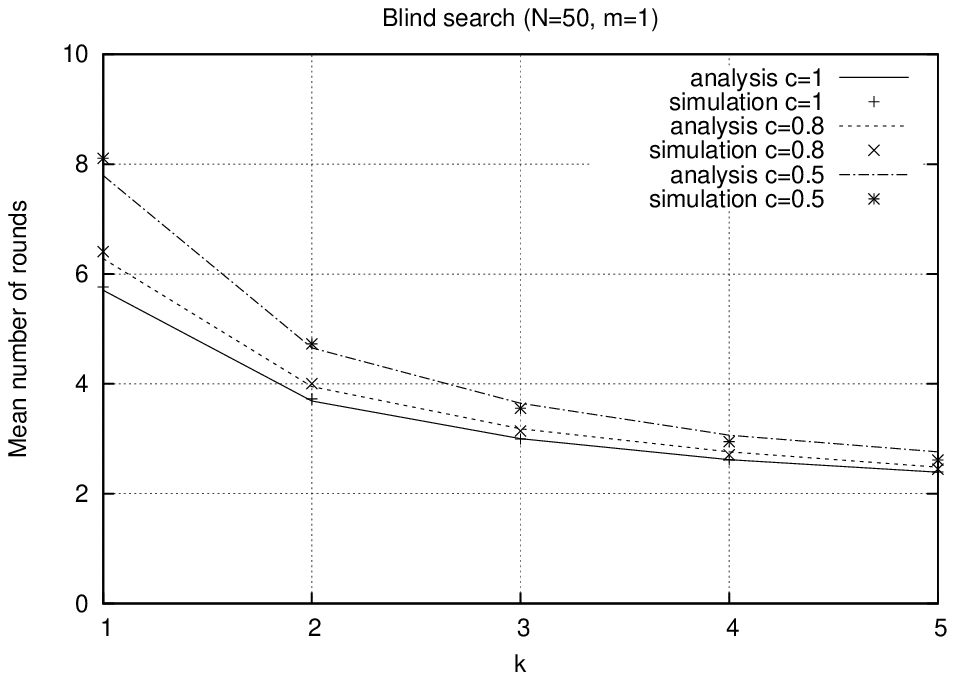}
\label{fig:subfig5}}%
\hspace{0pt}%
\subfigure[]{
\includegraphics[scale=0.75]{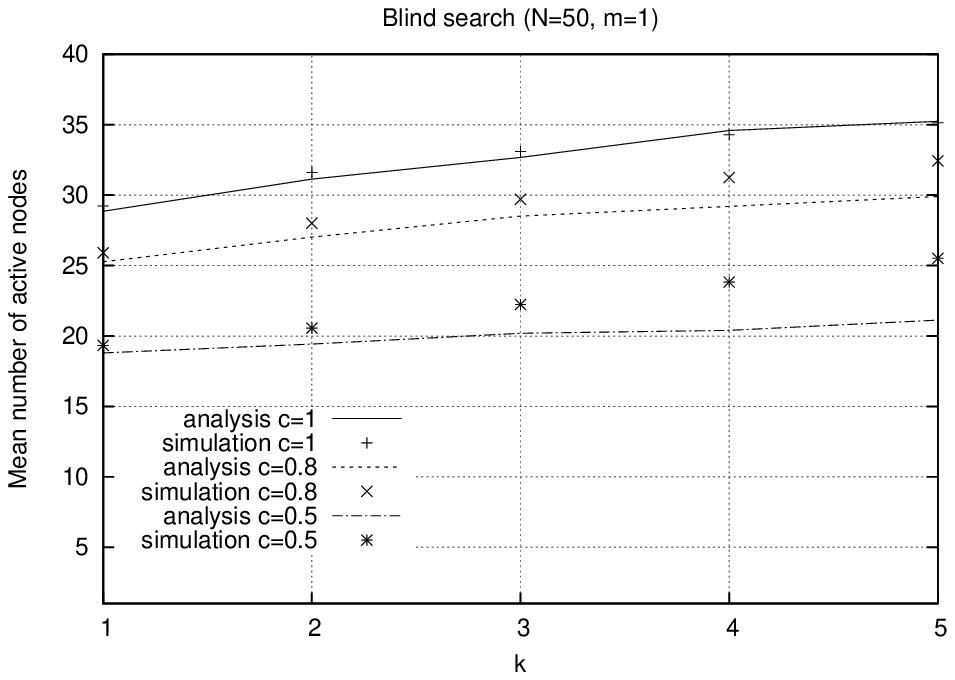}
\label{fig:subfig6}}
\caption[]{Analytical and simulation results for the mean number of rounds and mean number of active nodes
with varying network parameters, for the blind search algorithm\label{fig:bs_res}}
\end{figure}

We validate our approximation by means of simulation. The simulations were performed
with $100$ instances of random file positions, and $100$ random executions of
the search in each instance, leading to a total of $10^4$ repetitions in each experiment.
We evaluate the mean number of rounds and the mean number of activated
(infected) nodes until at least one copy of the file $f$ is found,
varying the number of nodes $N$ in the graph, the cooperation
probability $c$ and parameters $k$ and $m$. 
The value of $\epsilon$ in (\ref{eps}) was set to $10^{-6}$.
Results for different cases
are shown in Fig.~\ref{fig:bs_res}. 
\vspace{-0pt}

These figures illustrate that the simulation results match those from the theoretical analysis very well, 
except for large values of $k$.\footnote{It is noted that the improvement in accuracy 
by using the exact expression $(1-\frac{k}{N})^{A(r)}$,
or adding more terms in its expansion, is negligible.}
As the size of the network increases, the number of active nodes increases linearly,
while the increase in the mean number of rounds is superlinear, at a decreasing rate, as shown in Fig.~\ref{fig:subfig1} and \ref{fig:subfig2}. This
implies that the number of rounds can be well-fitted using a logarithmic function,
as will be shown more clearly in Section~\ref{sec:scaling}.

We examine the impact of search parameters $k$, $m$, in Fig.~\ref{fig:subfig3}-\ref{fig:subfig6}.
Note that the search can become faster
by increasing either $m$ or $k$. 
By comparing Fig.\ref{fig:subfig3} with
\ref{fig:subfig5}, we note that the increase in speed is higher for large values of $k$.
However, Fig.\ref{fig:subfig4} and \ref{fig:subfig6} show that increasing $k$ has the disadvantage
of increasing the number of active nodes, and thus produces higher communication overhead.
For example, calculations based on the simulation results show that for $c=1$ and $N=50$ nodes,
increasing $m$ to $3$ yields a relative decrease in the mean number of rounds
by $31\%$, and in the mean number of active nodes by $45\%$.
On the other hand, increasing $k$ to $3$ yields a higher relative decrease in
the mean number of rounds by $48\%$, but an {\em increase} in the mean number of
active nodes by $14\%$. Overall, we remark that increasing the number of queried neighbours results
in a great redundancy in the number of nodes that participate in the search with only small gains in speed.
\section{Analysis for the smart search case with cooperative or non-cooperative nodes}\label{sec:smart}
An analysis for the smart search process is presented below. An approximate analysis
similar to the one for the blind search model fails here, due to the varying probabilities
of successful query. Instead we adopt a direct combinatorial approach, by considering a
generalization of the occupancy (balls and bins) problem
\cite{Feller68}.

The generalized balls and bins problem is defined as follows: In a population of $n$ bins,
suppose we randomly distribute $r$ groups of $k$ balls, such that in
each group, no two balls go in the same bin and successive
distributions of groups of balls are independent. We want to find the
probability that exactly $v$ bins remain empty, where
$v=0,1,\dots,n-k$ (it is assumed that $n>k$).

We follow the approach in \cite[Section IV.2]{Feller68} for the
classical occupancy problem (where $k=1$). The total number of ways
to distribute $r$ groups of balls in the way described above is
$\binom{n}{k}^r$. Similarly, the total number of ways to assign them
to $n-1$ bins is $\binom{n-1}{k}^r$, so that the probability that
one {\em given} bin is empty, is
${\binom{n-1}{k}^r}/{\binom{n}{k}^r}$. Generally, the probability
that $v$ {\em given} bins are empty is
${\binom{n-v}{k}^r}/{\binom{n}{k}^r}$.

Therefore, the probability that at least one bin is empty is, by the
inclusion-exclusion method,
\begin{displaymath}
\sum_{i=1}^{n-k}{(-1)}^{i-1}\binom{n}{i}\frac{\binom{n-i}{k}^r}{\binom{n}{k}^r}\;.
\end{displaymath}
The probability that all bins are occupied, denoted by $p_0(r,k,n)$,
is
\begin{align}
p_0(r,k,n)&=1-\sum_{i=1}^{n-k}{(-1)}^{i-1}\binom{n}{i}\frac{\binom{n-i}{k}^r}{\binom{n}{k}^r}\nonumber\\
&=\sum_{i=0}^{n-k}{(-1)}^{i}\binom{n}{i}\frac{\binom{n-i}{k}^r}{\binom{n}{k}^r}\;.
\end{align}
Consider now the case where exactly $v$ non-given bins are empty. These $v$
bins can be chosen in $\binom{n}{v}$ different ways. The $k$ balls
of each of the $r$ groups are distributed among the remaining $n-v$
bins such that exactly $n-v$ are occupied. The mean number of such
distributions is
\begin{displaymath}
\binom{n-v}{k}^r p_0(r,k,n-v)\;.
\end{displaymath}
Dividing by the total number of possible configurations
$\binom{n}{k}^r$ we obtain the probability $p_v(r,k,n)$ that
exactly $v$ bins are empty:
\begin{equation}\label{gen_bb}
p_v(r,k,n)=\binom{n}{v}\frac{\binom{n-v}{k}^r}{\binom{n}{k}^r}
\sum_{i=0}^{n-k-v}{(-1)}^{i}\binom{n-v}{i}\frac{\binom{n-v-i}{k}^r}{\binom{n-v}{k}^r}\;.
\end{equation}

Based on (\ref{gen_bb}), we find
transition probabilities of the form $p(x_i,x_j)$, which denotes the
probability that if at a certain round of the algorithm there are
$x_i$ active nodes, then at the next round there will be $x_j$
active ones ($x_j\geq x_i$). It is emphasized here that each round corresponds
to one transition. In our terminology, {\em ``at (or in) a certain round''} will
have the meaning {\em ``after the transition that occured in this round
and before the next transition''}. The first round marks the transition from 1 active
node (the initiator) to a maximum number of $k+1$ active nodes.

Since there are no repetitions for the smart search, the transition
probabilities can be found by directly applying (\ref{gen_bb}),
substituting $n=N-x_i$, $v=N-x_j$, and $r=x_i$:
\begin{equation}\label{tr_pr_norep}
p(x_i,x_j)=p_{N-x_j}(x_i,k,N-x_i)\;.
\end{equation}

From (\ref{gen_bb}),(\ref{tr_pr_norep}) we have
\begin{align}
p(x_i,x_j)&=\binom{N-x_i}{N-x_j}\frac{\binom{x_j-x_i}{k}^{x_i}}{\binom{N-x_i}{k}^{x_i}}
\sum_{\ell=0}^{x_j-x_i-k}(-1)^{\ell}\binom{x_j-x_i}{\ell}\frac{\binom{x_j-x_i-\ell}{k}^{x_i}}{\binom{x_j-x_i}{k}^{x_i}}\;.
\end{align}

Each of the $x_j-x_i$ queried nodes will decide whether or not to be
cooperative independently with probability $c$. The probability
that $\alpha$ out of $x_j-x_i$ nodes will actually be activated is
\begin{displaymath}
B(x_j-x_i,\alpha,c)\stackrel{def}{=}\binom{x_j-x_i}{\alpha}c^{\alpha}(1-c)^{x_j-x_i-\alpha}\;.
\end{displaymath}

Therefore, the probability that there will be $x_i+\alpha$ active
nodes in the next round is
\begin{equation}
p(x_i,x_i+\alpha)=\sum_{x_j-x_i>\alpha} p(x_i,x_j)
B(x_j-x_i,\alpha,c)\;.
\end{equation}

Based on the transition probabilities, we construct the
$N\times N$ transition matrix $Q$ with entries $p(x_i,x_j)$
for $i,j=1,\dots,N$. The value of the $i$th element of the
first row of the matrix $Q^r$ is the probability that there are $i$
active nodes at round $r$.

Let us denote by $p_s(v)$ the probability that at least one of the $v$ active nodes
finds a copy of the file. The probability $S(r)$ of finding the file by (and including) round $r$ is
\begin{equation}\label{prob_findby_smart}
S(r)=\sum_{v}Q^{(r-1)}(1,v)\left(1-(1-p_{s}(v))^v\right)\;,
\end{equation}
where $p_{s}(v)$ is the probability that a search by a single node finds
a copy of the file, given that there are already $v$ active nodes.

To find $p_{s}(v)$, we take a similar approach as in Section~\ref{approx_blind_search}.
It is
\begin{equation}
p_{s}(v)=1-\frac{\binom{N-v-m}{k}}{\binom{N-v}{k}}
\end{equation}

Finally, the probability of finding the file at the $r$-th round is given by (\ref{prob_find_bs}).\footnote{
Notice that (\ref{prob_findby_smart}) is not a cdf, so we can't use $S(r)-S(r-1)$ to
find the probability of successful search at round $r$.}
We emphasize that this formula is again an approximation, since it is assumed that
each round is independent of the other. 

Based on the above distribution, we easily derive the expected
number of rounds until a file is found, as in (\ref{mean_rounds}).
The mean number of nodes activated during the search process is
\begin{equation}\label{mean_infected}
E[A]=\sum_{r=1}^\infty E[\alpha(r)]p(r)\;,
\end{equation}
where $E[\alpha(r)]$ is the mean number of active nodes in round
$r$, derived from the distribution $Q^r(1,\cdot)$. (For the smart search, the summation index in
(\ref{mean_rounds}),(\ref{mean_infected}) is upper bounded by
$\lceil{N-1}/k\rceil$.)

\begin{figure}[!htb]
\centering \subfigure[]{
\includegraphics[scale=0.75]{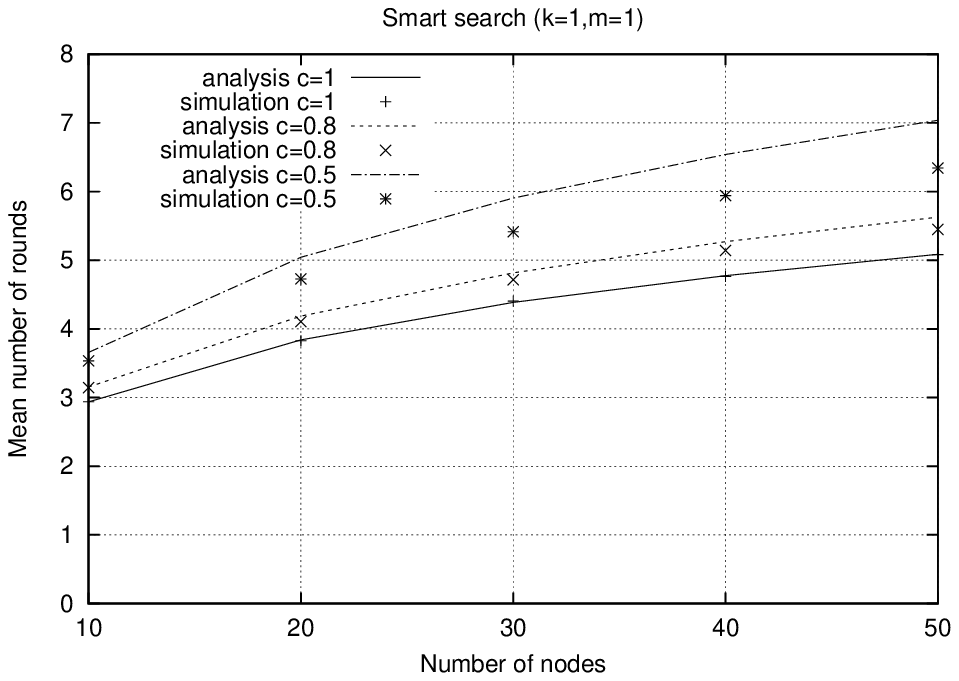}
\label{fig:subfig7}} %
\hspace{0pt}%
\subfigure[]{
\includegraphics[scale=0.75]{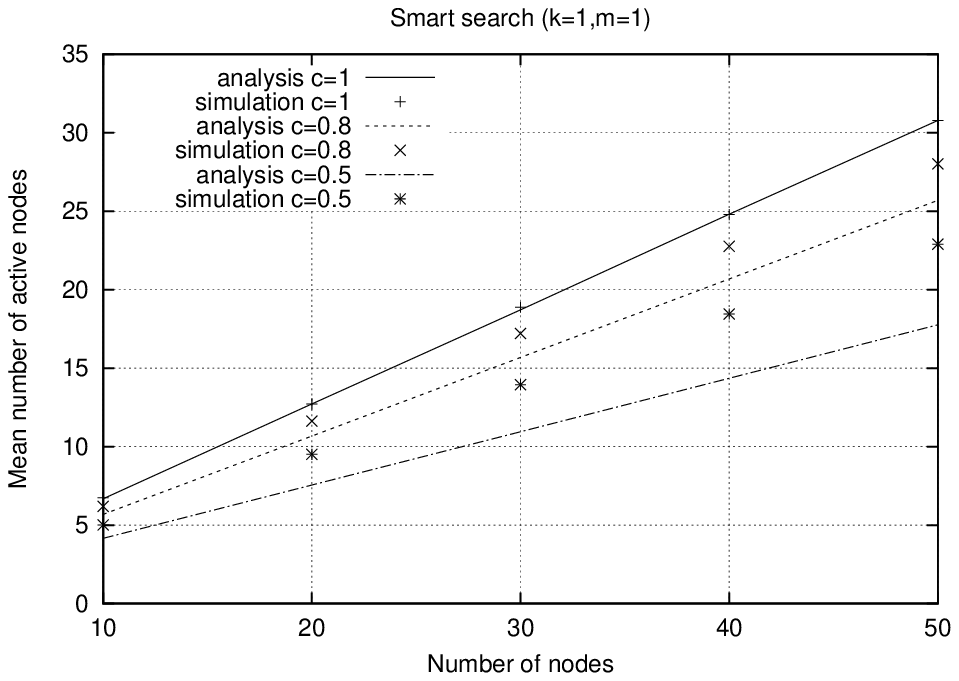}
\label{fig:subfig8}}\\%
\vspace{6pt}%
\subfigure[]{
\includegraphics[scale=0.75]{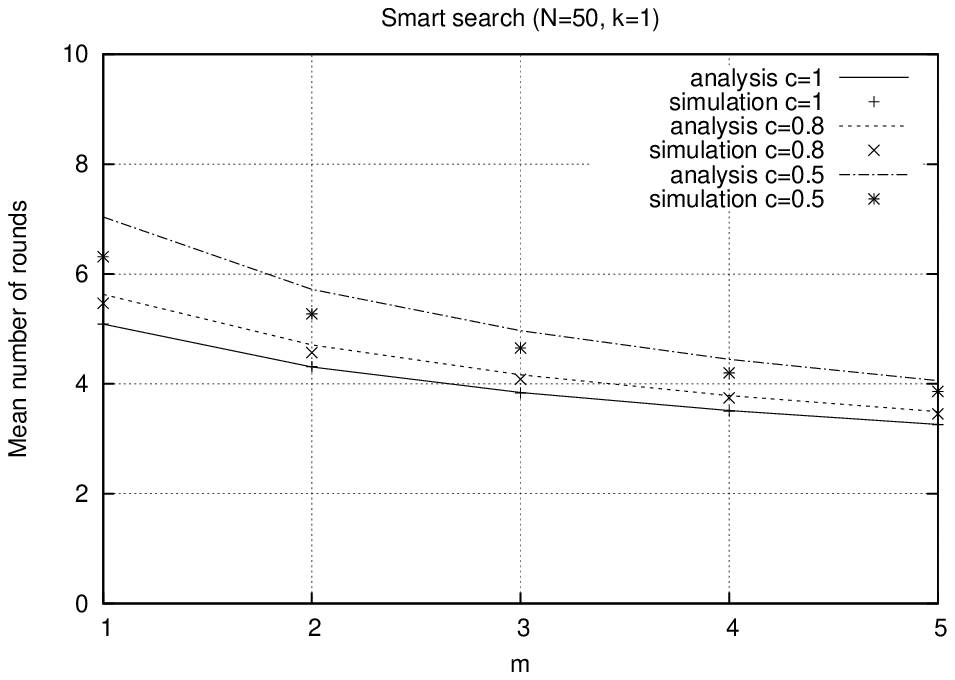}
\label{fig:subfig9}}%
\hspace{0pt}%
\subfigure[]{
\includegraphics[scale=0.75]{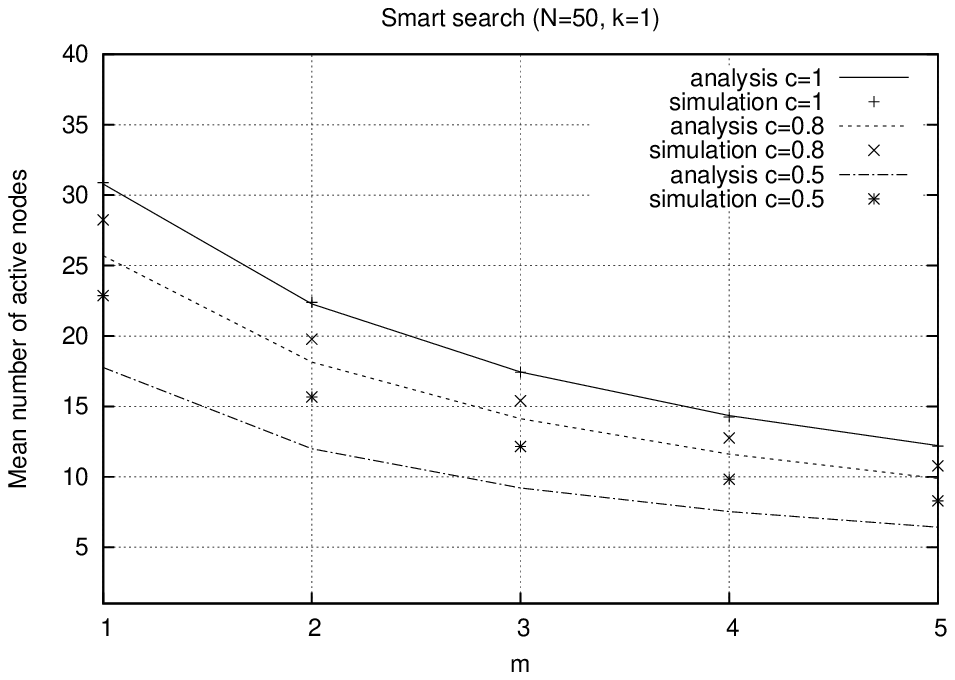}
\label{fig:subfig10}}\\%
\vspace{6pt}%
\subfigure[]{
\includegraphics[scale=0.75]{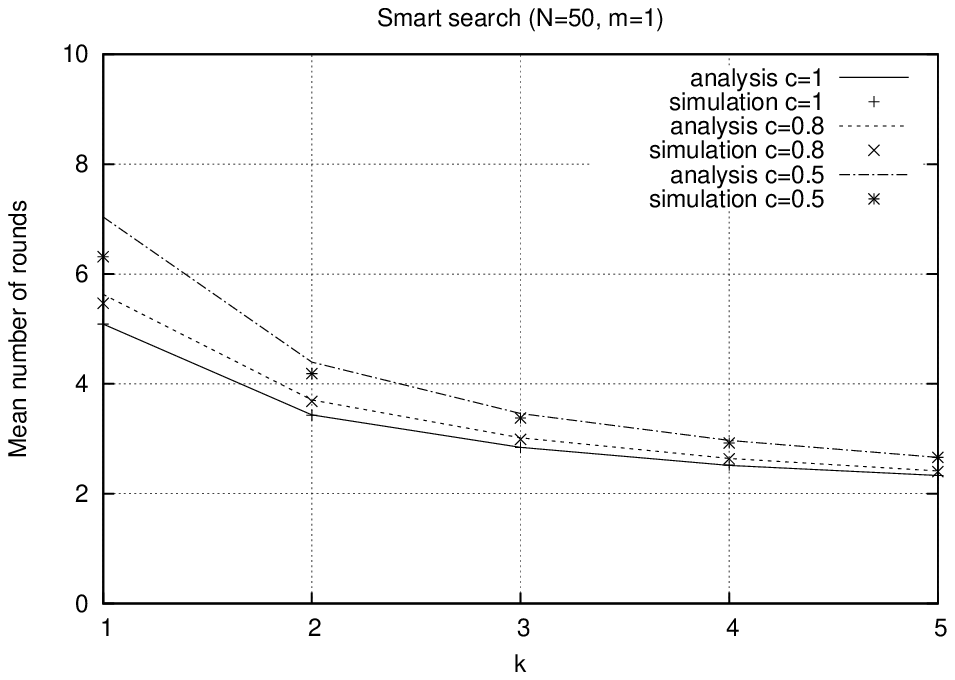}
\label{fig:subfig11}}%
\hspace{0pt}%
\subfigure[]{
\includegraphics[scale=0.75]{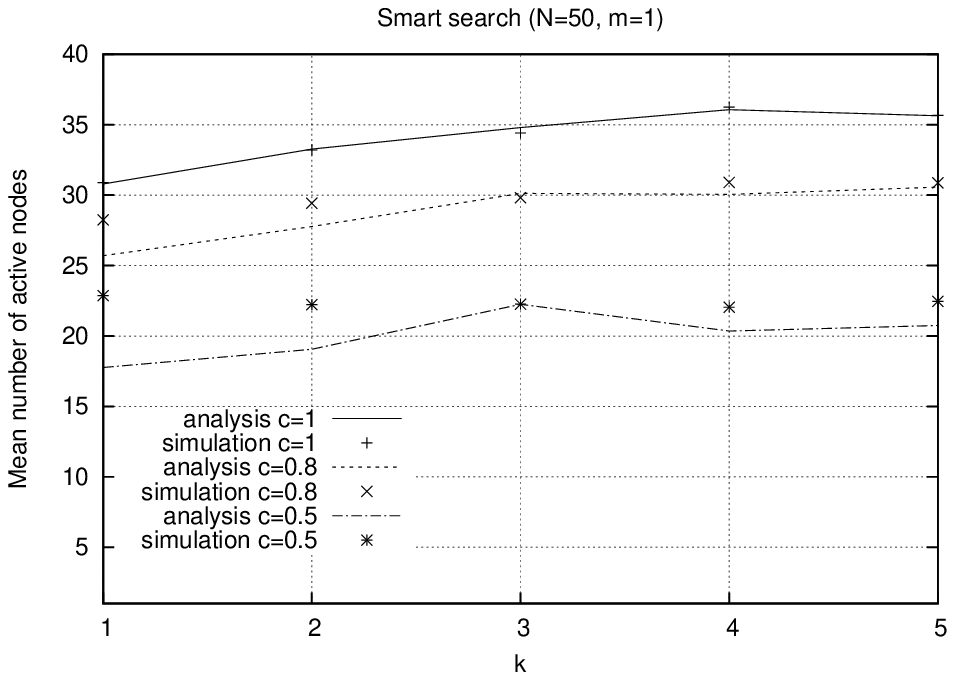}
\label{fig:subfig12}}
\caption[]{Analytical and simulation results for the mean number of rounds and mean number of active nodes
with varying network parameters, for the smart search algorithm\label{fig:smart_res}}
\end{figure}

We take both analytical and simulation results for the same values of the parameters $N$,
$c$, $k$ and $m$ as for the blind search algorithm. The simulations are conducted for the
same number of repetitions as for blind search. Results are
shown in Fig.~\ref{fig:smart_res}. Generally, the model is extremely accurate
for $c=1$, but as the cooperation probability decreases it starts to
deviate from the simulated behavior. For small values of $c$, we remark that the
model is less accurate than our model for blind search, even though it follows
a combinatorial approach that is exact up to (\ref{prob_findby_smart}). We attribute this to the fact that
the intermediate search steps are more correlated for the smart search
algorithm.
Thus, as it is intuitively reasonable, the assumption of independence over rounds leads to worse results for the smart search than
for the blind search case.

The same observations hold regarding the effect of the cooperation probability, the number of
queried neighbours and the number of copies of the file, as in the blind search case. 
It is again emphasized that the behavior with respect to increasing $k$ is an outcome 
of the tradeoff between the increased speed of discovery and the redundancy in the total number of messages sent.

We notice that there is only a small performance improvement of the smart
over the blind search algorithm, expressed through the decrease in the mean
number of rounds. This improvement becomes less pronounced as $k$ or
$m$ increase. For example, based on the simulation results for $N=50$ and
$c=1$, the relative reduction of smart search in the mean number of rounds
is $13\%$ when $k=1,\,m=1$, $3\%$ when $k=1,\,m=3$, and $5\%$ when $k=3,\,m=1$.
This impovement is relatively larger when the cooperation probability decreases:
for $N=50$ and $c=0.5$, the corresponding reductions were $27\%$, $16\%$ and $6\%$.

However, the mean number of active nodes may be greater for the smart search,
due to the fact that we always query only inactive nodes. This was true in most of
the derived results. For $N=50$ nodes, the relative increased reached up to $10\%$
for $c=1$ and $k=1,\,m=3$, while for $c=0.5$ it increased up to $31\%$ for $k=1,\,m=3$.
For the values of $N=50$ and $c=0.5$, only a slight decrease of $1\%$ was observed when $k=3,\,m=1$.

The overall results illustrate that when comparing the two cases, the smart
search does not offer a significant improvement. This leads us to the conclusion that if the overhead incurred
in the smart search algorithm for informing all active nodes of the identities of
queried nodes is not negligible compared to that of the search procedure, it is highly likely
that there is not much to gain by such a scheme.
\section{Analysis with stiflers}\label{sec:stiflers}
Another behavioral pattern that we consider is {\em stifling}.
In this pattern, each of the nodes that are (or become) active at a
certain round may cease to be active and not participate in the
search process any more. This could express a node's loss of
interest in spreading the query message further in the network.

We analyse this stifling behaviour based on the assumption that
at each round of the search, each active node may become a stifler
independently with probability $s$. A node that becomes a stifler is
considered as inactive, and in a blind search it may become
active again with probability $1-s$, if queried. We consider that the
initiator does not become a stifler, so the number of active nodes will
always be greater than zero.

This stifling behaviour will be modeled only for the blind search case, based on
our approximative method. In order to model the smart search case,
one has to discriminate between active and queried nodes, which leads to a
multi-dimensional Markov chain which is not easily amenable to analysis.

Given that there are $i(r)$ inactive nodes at round $r$, we are interested in
finding the mean number of inactive nodes at round $r+1$.
This consists of the mean number of active
nodes at round $r$ that became inactive (excluding the initiator) and the mean number of inactive nodes at round $r$ that remained inactive.
Hence,
\begin{align}
E[I(r+1)]=&(A(r)-1)s+I(r)\left[\left(1-\frac{k}{N-1}\right)^{{A}(r)}+\left[1-\left(1-\frac{k}{N-1}\right)^{{A}(r)}\right]s\right]\nonumber\\
=&(A(r)-1)s+i(r)\left[s+(1-s)\left(1-\frac{k}{N-1}\right)^{{A}(r)}\right]\nonumber\;.
\end{align}

Assuming $I(r)=i(r)\:\forall r$, and using again that $(1-\frac{k}{N-1})^{{A}(r)}\approx e^{-{A}(r)(\frac{k}{N-1}+\frac{k^2}{2(N-1)^2})}$, $A(r)=N-I(r)$,
we finally obtain the deterministic approximation
\begin{equation}
\begin{split}
A(r+1)=1+(N-1)(1-s)-(N-A(r))(1-s)e^{-{A}(r)(\frac{k}{N-1}+\frac{k^2}{2(N-1)^2})}\;,
\end{split}
\end{equation}
with $A(1)=1$.

From there we can follow similar steps as in Section~\ref{approx_blind_search} to find performance
measures of interest.
%\subsection{Results}

\begin{figure}[!tb]
\centering
\subfigure[]{
\includegraphics[scale=0.75]{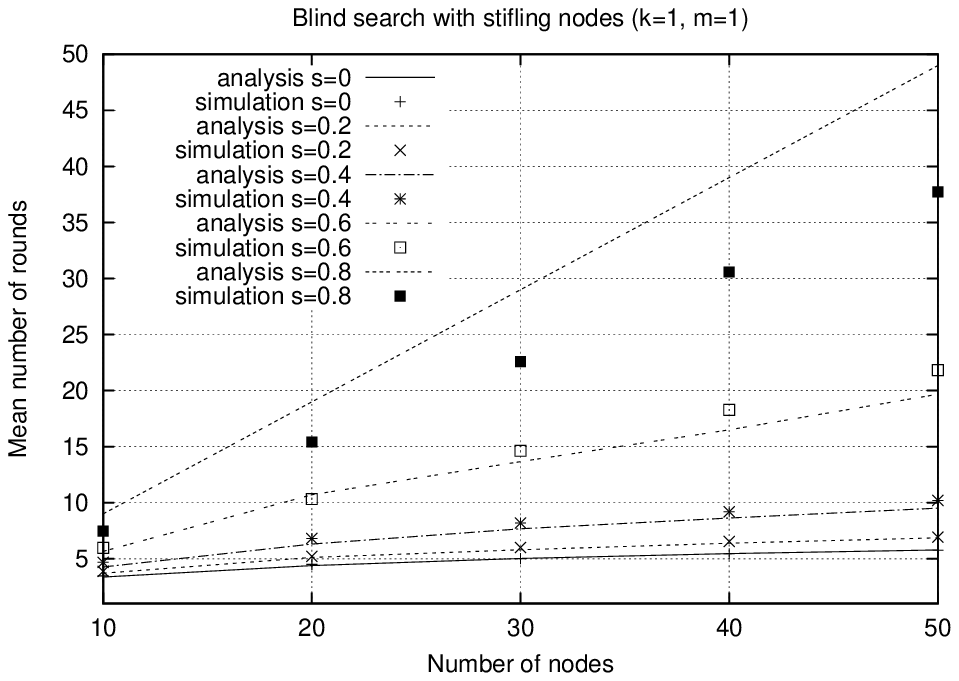}
\label{fig:subfig13}} %
\hspace{0pt}%
\subfigure[]{
\includegraphics[scale=0.75]{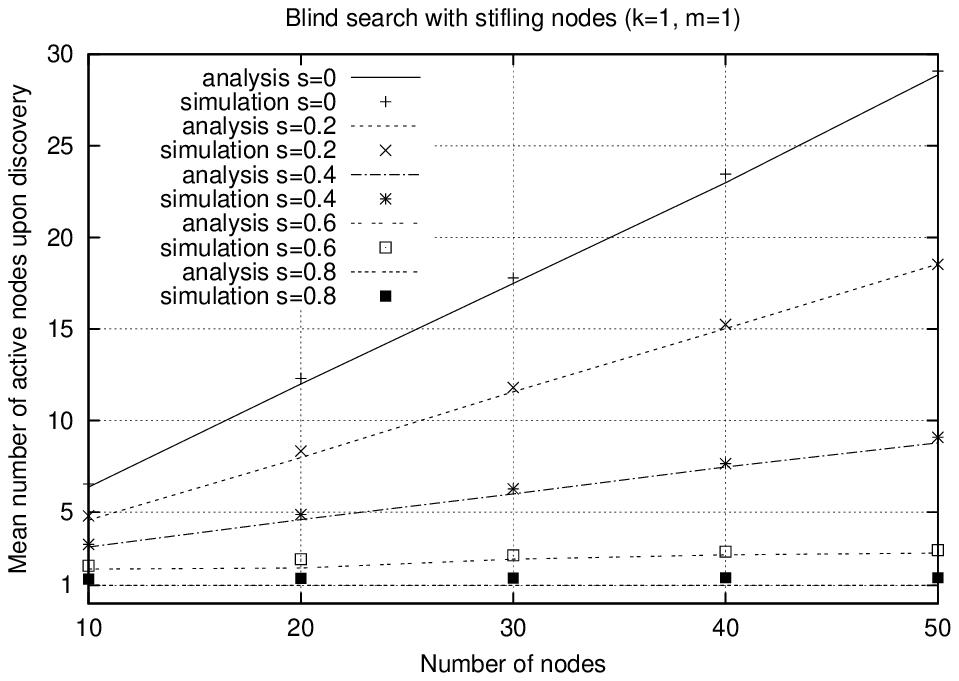}
\label{fig:subfig14}}\\%
%\vspace{6pt}%
%\subfigure[]{
%\includegraphics[scale=0.75]{results/blind_search_stiflers_k1m3_rounds.eps}
%\label{fig:subfig15}}%
%\hspace{0pt}%
%\subfigure[]{
%\includegraphics[scale=0.75]{results/blind_search_stiflers_k1m3_active.eps}
%\label{fig:subfig16}}\\%
%\vspace{6pt}%
%\subfigure[]{
%\includegraphics[scale=0.75]{results/blind_search_stiflers_k3m1_rounds.eps}
%\label{fig:subfig17}}%
%\hspace{0pt}%
%\subfigure[]{
%\includegraphics[scale=0.75]{results/blind_search_stiflers_k3m1_active.eps}
%\label{fig:subfig18}}
%%
\caption[]{Mean number of rounds and mean number of active nodes upon discovery
with $k=1$, $m=1$ for the blind search algorithm with stiflers
\label{fig:bs_stiflers_res}}
\end{figure}

Results based on simulation and the analytical approximation are shown in Fig.~\ref{fig:bs_stiflers_res}, 
for different values of parameters $m$, $k$, and the stifling probability $s$.
As $s$ increases, the performance of the search algorithm deteriorates.

We observe that the approximate model follows very well the simulated behaviour, except for
larger deviations in the mean number of rounds when the stifling probability is high.
This is mainly due to the higher relative error that results from the rounding operation
(see the analysis of Section~\ref{approx_blind_search}).
As the stifling probability gets higher, the number of active nodes in the network is rounded to one in the model,
and therefore the mean number of rounds approaches the inverse of the probability of a successful query of this node
(geometric distribution). For example, when $k=1$, the mean number of rounds approaches the
value of $(N-1)/m$.

A comparison of the speed of blind search between the stifling and plain non-cooperative case
shows that the search algorithm performs worse in the presence of stifler nodes.
We may remark from the results that the relative increase in the number of rounds when nodes
behave as stiflers -- rather than as plain non-cooperative nodes -- becomes greater when the number
of nodes in the network increase, or when the stifling probability increases.
Since stifling is opposite to cooperation, it makes sense to examine
dual values of $s$, $c$, i.e. such that $s=1-c$ holds.
For $k=1\,,m=1$, and $N=50$, the mean number of rounds is increased by $9\%$ for nodes that behave as stiflers
with probability $s=0.2$, compared to the case of plain non-cooperative nodes with $c=0.8$.
The relative increase is $5\%$ when $N=10$. When $s=c=0.5$, the corresponding relative increase
is much greater, and amounts to $78\%$.

This difference becomes smaller as the search becomes faster,
i.e. when increasing either the $k$ or $m$ parameters.
Regarding the relative influence of the parameters $k$, $m$ to the efficiency of the search
the same observations hold, as in all previous cases.

The mean number of active nodes calculated here is the mean number of nodes that are active upon
discovery of the file. We therefore do not count nodes which were previously active in the search.
Hence, it should be noted that the mean number of active nodes displayed here is
only indicative of the communication overhead, as it does not count the nodes that were active in
intermediate rounds of the algorithm, and hence the corresponding communication costs.
Generally, our findings show that this number is much smaller when compared to the plain non-cooperative case,
where active nodes remain in that state until the end. For $k=1\,,m=1$, and $N=50$, the number of active
nodes upon discovery is decreased by $39\%$ in the stifling case with $s=0.2$, compared to the
plain non-cooperative case with $c=0.8$.

\begin{figure}[!tb]
\centering
\subfigure[]{
\includegraphics[scale=0.75]{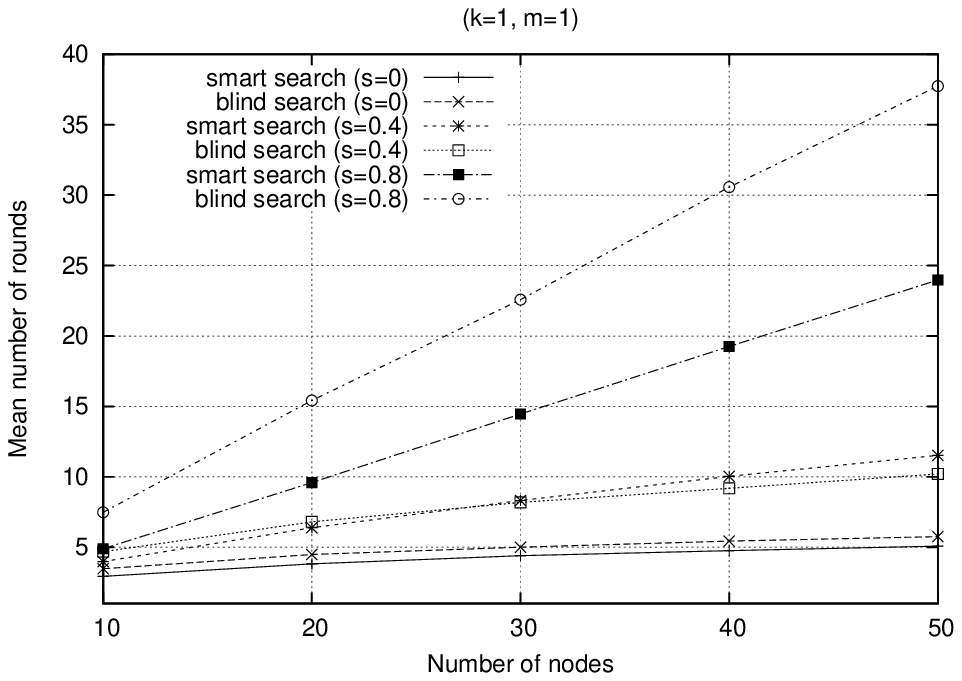}
\label{fig:bs_ss_k1m1_r}} %
\hspace{0pt}%
\subfigure[]{
\includegraphics[scale=0.75]{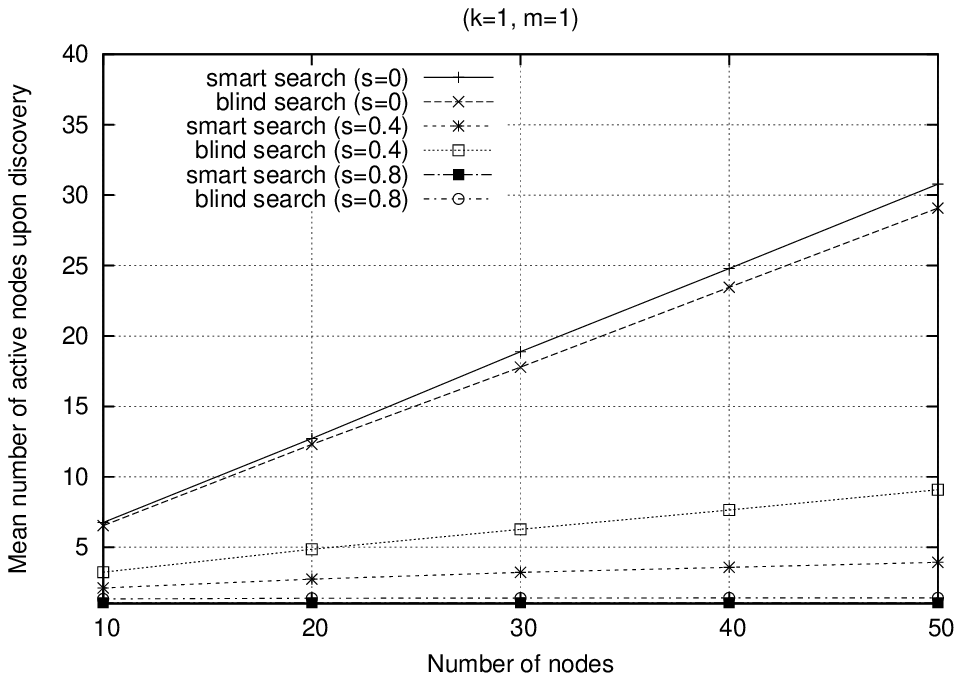}
\label{fig:bs_ss_k1m1_a}}\\%
\caption[]{Comparison of smart search against blind search with stiflers for $k=1\,,m=1$,
with different values of the stifling probability and increasing number of nodes.
\label{fig:bs_ss_k1m1}}
\end{figure}

We have also conducted a series of simulations to see the performance of smart search in the
presence of stifling nodes.
The smart search algorithm only queries nodes that have not been queried in previous rounds,
although all active nodes may become stiflers at any round.
In Fig.~\ref{fig:bs_ss_k1m1},
we show a comparison of smart search against blind search for $k=1\,,m=1$,
with different values of the stifling probability and increasing number of nodes.

We observe that smart search can yield a reduction in the number of rounds to discover the file,
which becomes significant for high values of the stifling probability.
For $s=0.8$ and $N=50$, the relative reduction is $37\%$.
However, the interesting thing is that for smaller values it may also yield a slight increase
(see the curves in Fig.~\ref{fig:bs_ss_k1m1_r} for $s=0.4$).
This seemingly unorthodox result is explained by the fact that since previously queried nodes are
not queried again, once they become stiflers, they permanently remain in that state and do not participate again in the search.
Thus trying to implement a ``smarter'' algorithm may also result in reducing the effective number of searchers,
thereby slowing down the search.

Fig.~\ref{fig:bs_ss_k1m1_a} shows that the mean number of active nodes upon discovery
is smaller for smart search than for blind search, when $s>0$ (as opposed to the plain non-coopera\-tive case, cf. Fig.~\ref{fig:bs_res},\ref{fig:smart_res}).
The relative decrease becomes greater for medium values of $s$ ($s=0.4$ in Fig.~\ref{fig:bs_ss_k1m1_a}). 
For high values of this probability, the mean number of active nodes approaches one and differences
become negligible. On the other hand, from Fig.~\ref{fig:bs_ss_k1m1_r} the highest gains in speed occur for
$s=0.8$. Therefore the highest gain in speed does not imply the highest reduction in redundant
active nodes, and vice-versa.
\section{Scaling performance of blind search}\label{sec:scaling}
\begin{figure}[!htb]
\centering
\subfigure[]{
\includegraphics[scale=1]{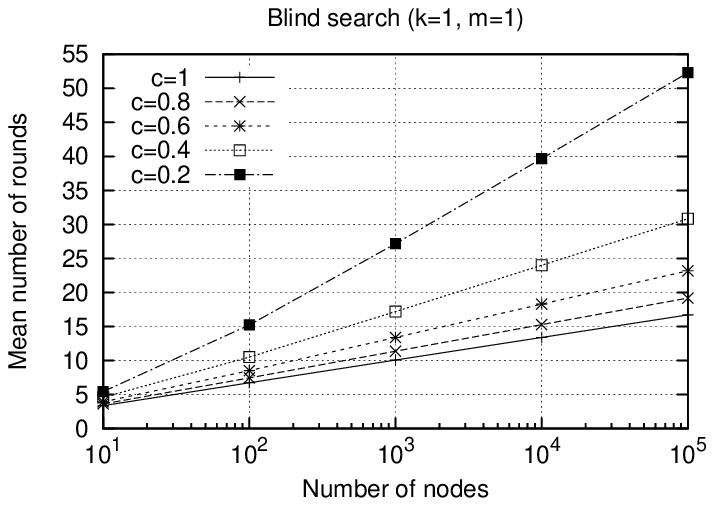}
\label{fig:large1}} %
\hspace{0pt}%
\subfigure[]{
\includegraphics[scale=1]{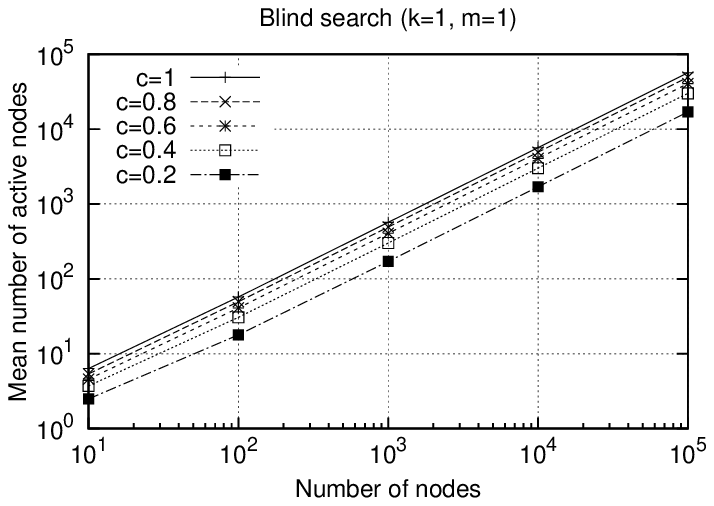}
\label{fig:large2}}\\%
\caption[]{Scaling performance of blind search in the plain non-cooperative case, for  $k=1$, $m=1$: (a) mean number of rounds,
(b) mean number of active nodes
\label{fig:bs_large_res}}
\end{figure}

\begin{figure}[!htb]
\centering
\subfigure[]{
\includegraphics[scale=1]{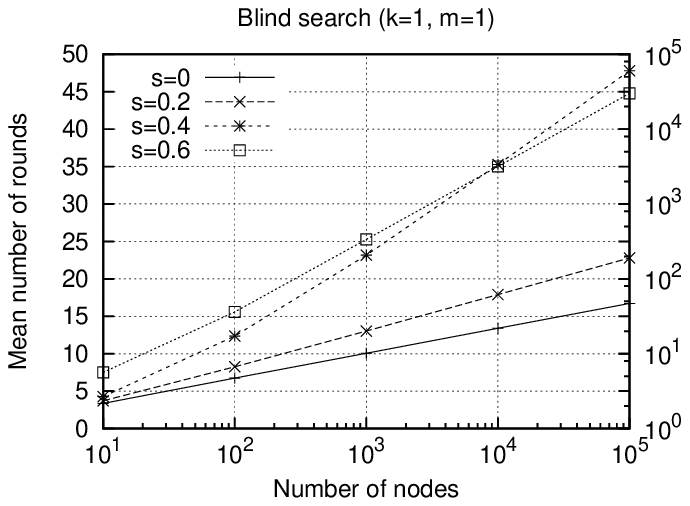}
\label{fig:large3}} %
\hspace{0pt}%
\subfigure[]{
\includegraphics[scale=1]{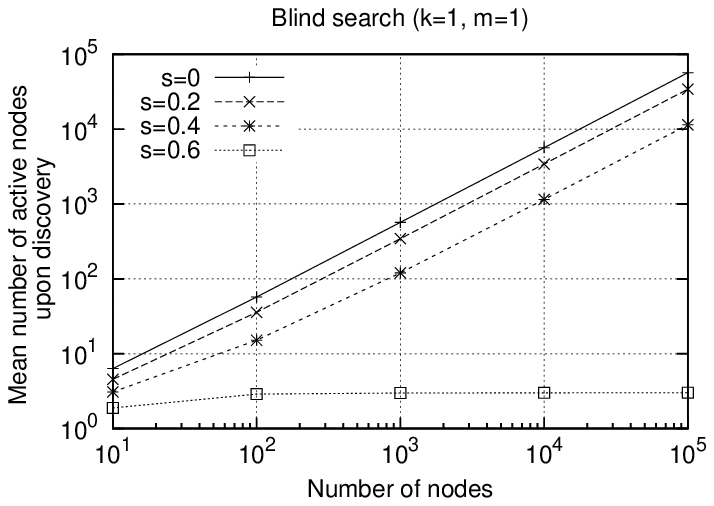}
\label{fig:large4}}\\%
\caption[]{Scaling performance of blind search in the stifling case, for  $k=1$, $m=1$: (a) mean number of rounds,
(b) mean number of active nodes upon discovery
\label{fig:bs_large_stiflers_res}}
\end{figure}

The low-complexity approximate model we have developed for the blind search algorithm enables
us to study its performance for networks with very large numbers of nodes.
We have taken results for networks with up to $10^5$ nodes, for both behavioural profiles:
plain non-cooperative and stifling. In Fig.~\ref{fig:bs_large_res},
we plot the mean number of rounds and the mean number of active nodes as a function of $N$
for the case of plain non-cooperative nodes, while in Fig.~\ref{fig:bs_large_stiflers_res},
similar results are taken for the case of stifling nodes.

The $x$-axis in all plots is in $\log$ scale. Fig.~\ref{fig:large2} and \ref{fig:large4}
are in log-log scale. In Fig.~\ref{fig:large3}, the curve for $s=0.6$ is scaled with respect
to the right $y$-axis.

From these results, we observe that the scaling performance of blind search is remarkably simple.
In the plain-non cooperative case, the mean number of rounds increases linearly with $\log N$,
while the mean number of active nodes increases linearly with $N$. This is true for almost the
whole range of values of $c$. (A more accurate estimate would be to consider a piece-wise linear function,
with a slightly smaller slope for $N<100$.)
In the case of stifling nodes, the behaviour is similar for values of $s<0.5$ approximately.
For large values of $s$, the mean number of rounds increases proportionately to the increase in
the number of nodes, while the mean number of active nodes upon discovery approaches 1.

Based on the observed behaviour, we can easily derive fitted functions,
based on the least squares method. For example, for $k=1\,,m=1$ and $c=1$, $E[r]=0.629\log N+0.057$,
and $E[A]=0.567N+0.584$.
\section{Conclusions}\label{sec:conclusions}
This paper has focused on the mathematical modeling of the gossip-based search algorithm in a complete graph.
We focused mainly on the blind-search algorithm, which is totally ignorant of previous queries and
can be implemented very easily, and compared its performance with respect to a smart-search algorithm,
where previously queried nodes are avoided.

Several conclusions can be extracted concerning trade-offs between speed, cost, and redundancy
between ``blind'' and ``smart'' gossip-based search algorithms.
We have investigated two extreme cases; many intermediate algorithms can be studied with
different levels of knowledge of previous queries, trading speed of discovery with additional
communication and processing overhead. An important observation is that speed
also trades-off with the redundancy in the number of query messages needed to locate a file.
The results we obtained provided serious indication that, when nodes have a plain non-cooperative
profile, the additional overhead of designing a ``smarter'' algorithm is not worth it,
since apart from the additional communication and processing cost, it induces high redundancy in the number of messages in the network.

For both the blind and smart search cases, we showed that the mean number of active nodes and the mean number of rounds
roughly increase linearly with the number of nodes in the network and its logarithm, respectively.
For the blind search algorithm, we were able to confirm this behaviour for very large numbers of nodes,
using the approximate model which has very low complexity. 

Another important observation concerns the relative impact on the search of the number of queried peers
by each node, and of the number of copies of the data object in the network. The increase of both these parameters
increases the speed of discovery. The relative increase is greater when the number of queried peers increases,
in both the blind and smart search algorithms. However, the corresponding increase in the number of active nodes
that (in most cases) ensues is inappropriate, and hence it is preferable to keep the value of this parameter very small.
The gossip-based search algorithm performs better when the requested data object is spread to many
nodes in the network.

Other useful remarks from this research concern the effects of different behavioural profiles: cooperative,
plain-non cooperative and stifling, with different degrees of cooperation. Stifling has a greater negative impact on
the search performance, which becomes worse in large networks.

Future research issues we envisage are mostly related to the performance evaluation of the gossip-based search algorithm.
The most significant direction of research is to examine the efficiency of the search in different types of networks.
From the network in the form of a complete graph that we studied here, we can pass to more general graphs
that are often met, such as Erd\"os-R\'enyi graphs, or graphs with power-law degree distribution (scale-free networks).
Performance results in such networks where each node has connections with different peers will give more realistic
evidence of its efficiency and applicability. Finally, it is interesting to compare the performance of the
algorithm with different distributed search schemes, in terms of speed and implementation cost.
\appendix
\section{Comparison between the approximate and the exact model for the blind search algorithm} 
We compare the approximate model for the blind search algorithm with the exact model developed in \cite{Tang09},
when the cooperation probability $c=1$. The two models are compared from the points of view of complexity and accuracy.
\subsection{Comparison of complexity}
We will compare the complexity of the two models based on the computational cost for deriving the probability
of locating the file at a certain round $r$, given by (\ref{prob_find_bs}) in the approximate and by (\ref{prob_findby_exact}) in the exact model.
The computational cost is measured based on the number of elementary steps needed to derive
the location probability, where each step consists of a small number of elementary operations
(addition, subtraction, multiplication or division).
We use the $O$-notation as the asymptotic upper bound of the complexity.

To compute (\ref{prob_find_bs}), first computations of (\ref{single-search}) and (\ref{incl-excl}) need to be
done. 
Equation (\ref{single-search}) involves two binomial coefficients,
which can be computed in $O(kN)$ time using the well-known linear recursion formula.\footnote{For $j\geq i> 0$,
$\binom{j}{i}=\binom{j-1}{i}+\binom{j-1}{i-1}$, with $\binom{j}{0}=\binom{j}{j}=1$.}
The recursive formula (\ref{approx_recursion}) involves a small number of multiplications and additions, and an
exponential function. 
The exponential can be calculated easily by splitting the exponent into integer and fractional parts (the latter
can be computed within high accuracy with a few terms only in a Taylor expansion, see e.g. \cite{Muller97}).
Hence each execution of the recursion has order one, and computing $A(r)$ or $\hat{A}(r)$ takes time
$O(r)$.
It holds that $\hat{A}(r)\leq N$. Therefore, the computation of (\ref{incl-excl}) and (\ref{prob_find_bs}),
given their input parameters, takes time $O(N)$ and $O(r)$
respectively, since in the first case it involves the computation of
$\hat{A}(r)$ polynomials, and in the second of $r$ polynomials, both
of degree one.
Therefore, the total complexity of deriving the probability of locating the file at round $r$, using the
approximate model is $O(kN+r)$.

To find the probability to find at least one copy of the file with the exact modeling,
we need to calculate the $r$-th power of the transition matrix $Q$ and then
solve equations (\ref{prob_findby_exact}),(\ref{prob_find_exact}) sequentially.
The first computation involves the multiplication of an $N\times N$ transition probability
matrix, with complexity at worst $O\left(  N^{3}\right)$.
For sufficiently large $r$, we can consider the sequence of matrices $Q$, $Q^{2}$,
$Q^{4}$, $...$, $Q^{2^{k}}$, instead of computing the sequence $Q$, $Q^{2}$,
$...$, $Q^{r}$. Since the former one converges considerably faster compared
with the latter one. Therefore, we can compute the matrix power in $O\left(  \ln
(r)N^{3}\right)  $ steps.
The computation of (\ref{prob_findby_exact}) involves two binomial coefficients, namely,
$\binom{N-i}{m}$ and $\binom{N-1}{m}$,
which have complexity $O\left(  mN\right)$ . Therefore, it takes
$O\left(mN^2\right) $ steps to solve (\ref{prob_findby_exact}). The total computational complexity
is thus dominated by the complexity to compute the matrix power, which is $O\left(\ln
(r)N^{3}\right)$ in our case.
\subsection{Comparison of accuracy}
We next compare the relative accuracy of the approximate model for calculating
the mean number of steps to find at least one copy of the file
and the mean number of nodes activated in the search.
The relative accuracy is calculated as $(1-|exact-approx|/exact)100\%$, and is output with two decimal digits.
It is reminded that the
comparison is done when the cooperation probability is one. Results are shown in Table~\ref{accur_table} below,
where $N$ stands for the total number of nodes in the network (including the initiator).
\vspace{-3pt}
\begin{table}[!htb]
\caption{Relative accuracy (\%) of the approximate model for blind search}
\small
\centering
\subtable[Mean number of rounds]{
\begin{tabular}
[c]{c|ccccc}%\hline1
 & $N=10$ & $N=20$ & $N=30$ & $N=40$ & $N=50$\\\hline
$k=1$, $m=1$ & $94.07$ & $97.23$ & $98.94$ & $99.60$ & $100$\\
$k=1$, $m=3$ & $93.97$ & $97.00$ & $98.45$ & $99.18$ & $99.75$\\
$k=3$, $m=1$ & $96.16$ & $98.77$ & $99.77$ & $99.89$ & $99.70$\\
\end{tabular}
}
\subtable[Mean number of nodes activated in the search]{
\centering
\begin{tabular}
[c]{c|ccccc}%\hline
 & $N=10$ & $N=20$ & $N=30$ & $N=40$ & $N=50$\\\hline
$k=1$, $m=1$ & $92.60$ & $95.78$ & $96.12$ & $96.34$ & $97.81$\\
$k=1$, $m=3$ & $95.43$ & $95.97$ & $94.81$ & $95.25$ & $96.58$\\
$k=3$, $m=1$ & $86.69$ & $90.51$ & $93.63$ & $94.98$ & $95.97$\\
\end{tabular}
}\label{accur_table}
\end{table}

These results confirm that the approximation becomes more accurate when the number of nodes
in the network increases. The greatest inaccuracy is observed for a relatively large -- compared to $N$ --
number of queried neighbours $k$, as was also indicated in Fig.~\ref{fig:subfig6}.
Generally, the model proves to be credible, with a relative accuracy that is higher than 95\% in
the majority of the above cases.

\begin{thebibliography}{1}

\bibitem{Eugster04}
P.~T. Eugster, R.~Guerraoui, A.-M. Kermarrec, and L.~Massouli\'e.
\newblock Epidemic information dissemination in distributed systems.
\newblock {\em IEEE Computer}, 37(5):60--67, 2004.

\bibitem{Feller68}
W.~Feller.
\newblock {\em An Introduction to Probability Theory and Its Applications,
  Volume 1}.
\newblock Wiley, January 1968.

\bibitem{Kempe03}
D.~Kempe, A.~Dobra, and J.~Gehrke.
\newblock Gossip-based computation of aggregate information.
\newblock {\em Foundations of Computer Science, Annual IEEE Symposium on},
  0:482, 2003.

\bibitem{Muller97}
J.-M. Muller.
\newblock {\em Elementary Functions, Algorithms and Implementation}.
\newblock Birkhauser, Boston, 1997.

\bibitem{Nekovee08}
M.~Nekovee, Y.~Moreno, G.~Bianconi, and M.~Marsili.
\newblock Theory of rumour spreading in complex social networks.
\newblock {\em Physica A}, 374:457, Jul 2008.

\bibitem{Pittel87}
B.~Pittel.
\newblock On spreading a rumor.
\newblock {\em SIAM J. Appl. Math.}, 47(1):213--223, 1987.

\bibitem{Spyropoulos08}
T.~Spyropoulos, K.~Psounis, and C.~S. Raghavendra.
\newblock Efficient routing in intermittently connected mobile networks: the
  single-copy case.
\newblock {\em IEEE/ACM Trans. Netw.}, 16(1):63--76, 2008.

\bibitem{Tang09}
S.~Tang, E.~Jaho, I.~Stavrakakis, I.~Koukoutsidis, and P.~V. Mieghem.
\newblock Modeling gossip-based content propagation and search in distributed
  p2p overlay.
\newblock {\em Computer Networks (under submission)}, 2009.

\end{thebibliography}
\end{document}